\documentclass[showkeys,pra,12pt,amsmath,amssymb,tightenlines]{revtex4}
\usepackage{graphicx}

\begin{document}

\title{Chebyshev polynomials and Fourier transform\\
of $SU(2)$ irreducible representation character\\
as spin-tomographic star-product kernel}

\author{S. N. Filippov$^1$}
\email{filippovsn@gmail.com}

\author{V. I. Man'ko$^2$}
\email{manko@sci.lebedev.ru}

\affiliation{$^1$Moscow Institute of Physics and Technology (State
University) \\Institutskii per. 9, Dolgoprudnyi, Moscow Region
141700, Russia
\\$^2$P. N. Lebedev Physical Institute, Russian Academy of Sciences \\Leninskii Prospect 53, Moscow 119991, Russia}

\begin{abstract}
Spin-tomographic symbols of qudit states and spin observables are
studied. Spin observables are associated with the functions on a
manifold whose points are labelled by spin projections and sphere
$S^{2}$ coordinates. The star-product kernel for such functions is
obtained in explicit form and connected with Fourier transform of
characters of $SU(2)$ irreducible representation. The kernels are
shown to be in close relation to the Chebyshev polynomials. Using
specific properties of these polynomials, we establish the
recurrence relation between kernels for different spins. Employing
the explicit form of the star-product kernel, a sum rule for
Clebsch-Gordan and Racah coefficients is derived. Explicit
formulas are obtained for the dual tomographic star-product kernel
as well as for intertwining kernels which relate spin-tomographic
symbols and dual tomographic symbols.
\end{abstract}


\maketitle

\section{\label{introduction}Introduction}

In quantum mechanics, states of a system are usually associated
with the density operators. The other possibility is to use
different maps of quantum states onto the quasi-probability
functions \cite{wigner,husimi} or the fair
probability-distribution function called tomogram (see, e.g.,
\cite{tombesi-manko,sudarshan-2008,berber,vogel,mendes-physica,oman'ko-97,marmoJPA}).
The latter one is of great interest because it can be measured
experimentally
\cite{raymer,mlynek,lvovsky,solimeno-porzio,bellini}. Though
tomograms are often utilized with the only aim to reconstruct the
Wigner function or the density matrix, it should be emphasized
that tomograms themselves are a primary notion of quantum states.
As far as spin states are concerned, the corresponding tomographic
map is elaborated in
\cite{dodonovPLA,oman'ko-jetp,weigert,amiet,agarwal}. The examples
of other maps of spin states onto functions are discussed in
\cite{klimov-sanchez-soto,garcia-bondia,klimov-romero}. By analogy
with the density operator, any other operator (observable) on a
Hilbert space can be mapped onto the function called tomographic
symbol of the operator. Such a scanning procedure is acomplished
with the help of the special dequantizer operator
\cite{marmoJPA,vitale::dual}. Using the quantizer operator
\cite{marmoJPA,vitale::dual}, one can reconstruct the operator in
question, i.e, there exists an inverse map of tomographic symbols
onto operators. Within the framework of the spin-tomographic
star-product procedure
\cite{manko:star:1,manko:star:2,omanko-star-brief}, one deals with
symbols instead of operators. In particular, the symbol of the
product of two operators is equal to the star-product of the
symbols corresponding to the separate operators. The main feature
of the star-product is that it is associative but noncommutative
in general. The star-product kernel is easily expressed in terms
of the dequantizer and quantizer operators. The explicit formula
of the kernel was presented previously with the help of
Clebsch-Gordan and Racah coefficients in the work \cite{castanos}
and specified for the low-spin states in
\cite{filipp-spin-tomography}.

\bigskip

The aim of this work is to obtain a new explicit formula of the
spin-tomographic star-product kernel in terms of $SU(2)$
irreducible representation character. As mentioned above, any
qudit state can be described by the spin-tomographic probability
introduced in \cite{dodonovPLA,oman'ko-jetp}, where the
reconstructed states are expressed in terms of such state
characteristics as the Wigner function or the density matrix. In
the work \cite{d'ariano:paini}, the discussion is presented how to
use such a tomographic probability-distribution in the other known
reconstruction procedure. Both the spin tomogram, which coincides
with that introduced in \cite{dodonovPLA,oman'ko-jetp}, and the
inversion formula, which provides the density operator of a spin
state by means of its spin tomogram, were given in
\cite{d'ariano:paini} in the compact exponential forms. On the
other hand, it was proved in \cite{filipp-spin-tomography} that
the exponential form of the inversion formula, the inversion
formula found in \cite{oman'ko-jetp}, and that presented in
another form in \cite{castanos} are all identical on the set of
spin tomograms. In view of this equivalency, one can use any form
of the inversion formula on an equal footing. This means that any
form of the quantizer and dequantizer operators is acceptable to
study concrete properties of the star-product representation of
spin operators and qudit states. In the present paper, the problem
is attacked with the help of exponential representation of the
quantizer and dequantizer operators. The $SU(2)$ irreducible
representation character is known to be nothing else but the
Chebyshev polynomial of a specific argument. In turn, the kernel
is shown to be a Fourier transformation of the character. We
exploit special properties of the Chebyshev polynomials not only
to derive the star-product kernel but also to reveal its
peculiarities, for instance, the recurrence relation. Comparing
different explicit forms of the star-product kernels, we show that
the kernel is not defined unambiguously, but the residual must
give zero while integrating with tomographic symbols. We point out
that the constructed kernels of the spin-tomographic star-product
are given for the functions which depend not only on group element
of $SU(2)$ but also on weights (spin projection $m$) of $SU(2)$
irreps.

\bigskip

The paper is organized as follows.

In Sec. \ref{section:star:product:General}, we give a brief review
of the scanning and reconstruction procedures which are performed
in the spin tomography of qudit states. In Sec.
\ref{section:star:product:kernel}, the star-product kernel is
represented in the form of Fourier transformation of the $SU(2)$
irreps character and the explicit formula for the kernel is
derived. In Sec. \ref{section:equivalency}, we compare two
different explicit forms of the kernel to obtain a new sum rule
for Clebsch-Gordan and Racah coefficients. In Sec.
\ref{section:recurrence}, we establish the recurrence relation
between the star-product kernels for different spins. In Sec.
\ref{section:conclusions}, conclusions are presented. The
ambiguity of the spin-tomographic star-product kernel is
illustrated in Appendix \ref{appendix:equivalency}. Generalization
of the explicit formulas to other types of kernels is given in
Appendix \ref{appendix:generalization:of:star:product}.

\section{\label{section:star:product:General} Spin tomograms and tomographic symbols of operators}

Unless specifically stated, qudit states with spin $j$ are
considered. We deal with the angular momentum operators
$\hat{J}_{x}$, $\hat{J}_{y}$, and $\hat{J}_{z}$ and the standard
state vectors $|jm\rangle$ defined as follows:

\begin{equation}
\label{basis:states:j:m} (\hat{J}_{x}^{2} + \hat{J}_{y}^{2} +
\hat{J}_{z}^{2}) |jm\rangle = j(j+1) |jm\rangle, \qquad
\hat{J}_{z} |jm\rangle = m |jm\rangle,
\end{equation}

\noindent with the spin projection $m$ taking the values $-j,
-j+1, \dots ,j$.

As stated above, the state of a qudit is completely determined by
its density operator $\hat{\rho}$ or alternatively by the
following probability-distribution function (called spin
tomogram):

\begin{equation}
\label{tomogram} w_{j}(m, {\bf n}) = \langle jm |
\hat{R}^{\dag}({\bf n}) \hat{\rho} \hat{R}({\bf n}) | jm \rangle =
{\rm Tr} \Big( \hat{\rho} ~ \hat{R}({\bf n}) |j m \rangle \langle
jm | \hat{R}^{\dag}({\bf n}) \Big) = {\rm Tr} \Big( \hat{\rho}
\hat{U}_{j}(m, {\bf n}) \Big),
\end{equation}

\noindent where we introduced the dequantizer operator $\hat{U}(m,
{\bf n})$ and the rotation operator $\hat{R}({\bf n})$. The vector
${\bf n}(\theta,\phi) = (\cos\phi\sin\theta, \sin\phi\sin\theta,
\cos\theta)$ determines the axis of quantization (the point on the
sphere specified by the longitude $\phi \in [0,2\pi]$ and the
latitude $\theta \in [0,\pi]$). The rotation operator is defined
through

\begin{equation}
\hat{R}({\bf n}) = e^{- i ({\bf n}_{\bot} \cdot \ \hat{\bf J})
\theta }, \qquad {\bf n}_{\bot} = (-\sin\phi, \cos\phi, 0).
\end{equation}

%\exp \left[ i ({\bf n}_{\bot} \hat{\bf J}) {\theta}/{2}
%\right]

\noindent The tomogram satisfies the following normalization
conditions:

\begin{equation}
\sum \limits_{m=-j}^{j} w_{j} (m,{\bf n}) = 1,  \qquad
\frac{2j+1}{4\pi} \int \limits_{0}^{2\pi} d \phi \int
\limits_{0}^{\pi} \sin\theta d \theta \ w_{j} (m,{\bf
n}(\theta,\phi)) = 1.
\end{equation}

\noindent Some other features of the spin-tomographic functions
are discussed in
\cite{andreev:manko,safonov:manko,andreev:safonov:manko,filipp}.

Taking into account the relation $\hat{R}({\bf n}) \hat{J}_{z}
\hat{R}^{\dag}({\bf n}) = ({\bf n} \cdot \hat{\bf J}) \equiv
n_{\alpha} \hat{J}_{\alpha}$, the dequantizer operator
$\hat{U}_{j}(m, {\bf n})$ can be written in the following
exponential form:

\begin{eqnarray}
\label{dequantizer:exponential} \hat{U}_{j}(m, {\bf n}) && =
\hat{R}({\bf n}) | jm \rangle \langle jm | \hat{R}^{\dag}({\bf n})
= \hat{R}({\bf n}) \delta (m - \hat{J}_{z}) \hat{R}^{\dag}({\bf
n}) = \delta \left( m - \hat{R}({\bf n}) \hat{J}_{z}
\hat{R}^{\dag}({\bf n}) \right) \nonumber\\ && = \delta \left( m -
({\bf n} \cdot \hat{\bf J}) \right) = \frac{1}{2\pi} \int
\limits_{0}^{2\pi} e^{i m \varphi} e^{-i({\bf n} \cdot \hat{\bf
J}) \varphi} d \varphi,
\end{eqnarray}

\noindent where by $\delta$ we denote the Kronecker delta-symbol
\cite{mmanko}.

Given the tomogram (\ref{tomogram}), one can reconstruct the
density operator $\hat{\rho}$ with the help of the quantizer
operator $\hat{D}_{j}(m,{\bf n})$. The reconstruction procedure
reads

\begin{equation}
\hat{\rho} = \sum\limits_{m=-j}^{j} \frac{1}{4\pi}
\int\limits_{0}^{2\pi} d\phi \int\limits_{0}^{\pi} \sin \theta
d\theta ~ w_{j}(m,{\bf n}(\theta,\phi)) \hat{D}_{j}(m,{\bf
n}(\theta,\phi))
\end{equation}

\noindent or briefly

\begin{equation}
\label{rho:reconstruction} \hat{\rho} = \int w_{j}({\bf x})
\hat{D}_{j}({\bf x}) d {\bf x},
\end{equation}

\noindent where we denoted

\begin{equation}
{\bf x} = (m,{\bf n}), \qquad \int d {\bf x} =
\sum\limits_{m=-j}^{j} \frac{1}{4\pi} \int d\Omega =
\sum\limits_{m=-j}^{j} \frac{1}{4\pi} \int\limits_{0}^{2\pi} d\phi
\int\limits_{0}^{\pi} \sin \theta d\theta.
\end{equation}

Similarly to the case of dequantizer, the quantizer operator
$\hat{D}_{j}({\bf x})$ can be represented in the exponential form
\cite{d'ariano:paini}

\begin{equation}
\label{quantizer:exponential} \hat{D}_{j}(m, {\bf n}) =
\frac{2j+1}{\pi} \int \limits_{0}^{2\pi} \sin^{2} (\varphi / 2)
e^{i m \varphi} e^{- ({\bf n} \cdot \hat{\bf J}) \varphi} d
\varphi = \frac{2j+1}{2\pi} \sum\limits_{s=-1}^{1}
\frac{1}{1-3s^{2}} \int \limits_{0}^{2\pi} e^{i (m+s) \varphi}
e^{- i ({\bf n} \cdot \hat{\bf J}) \varphi} d \varphi.
\end{equation}

Both quantizer and dequantizer are Hermitian operators, with the
latter one being nonnegative as well.

By analogy with the density operator, any spin operator $\hat{A}$
acting on a Hilbert space of states (\ref{basis:states:j:m}) is
mapped onto the function $f_{\hat{A}}({\bf x})$ and vice versa. By
construction, the relation between the tomographic symbol
$f_{\hat{A}}({\bf x})$ and the operator $\hat{A}$ reads

\begin{equation}
f_{\hat{A}}({\bf x}) = {\rm Tr} \left( \hat{A}\hat{U}_{j}({\bf x})
\right), \qquad \hat{A} = \int f_{\hat{A}}({\bf x})
\hat{D}_{j}({\bf x}) d{\bf x}.
\end{equation}

Besides usual tomographic symbols described above, sometimes it is
convenient to use dual tomographic symbols defined through

\begin{equation}
f_{\hat{A}}^{d}({\bf x}) = {\rm Tr} \left( \hat{A}\hat{D}_{j}({\bf
x}) \right), \qquad \hat{A} = \int f_{\hat{A}}^{d}({\bf x})
\hat{U}_{j}({\bf x}) d{\bf x}.
\end{equation}

\noindent For instance, the average value of the operator
$\hat{A}$ reads

\begin{eqnarray}
{\rm Tr} \left(\hat{\rho}\hat{A}\right) = {\rm Tr} \int
w_{j}({\bf{x}}) \hat{D}_{j}({\bf{x}}) \hat{A} \ d {\bf{x}} = \int
w_{j}({\bf{x}}) {\rm Tr} \left( \hat{A} \hat{D}_{j}({\bf{x}})
\right) d {\bf{x}} = \int w_{j}({\bf{x}})
f_{\hat{A}}^{d}({\bf{x}}) d {\bf{x}}.
\end{eqnarray}

\noindent This implies that one can calculate mean values of
observables by using standard and dual tomographic symbols only.
The dual tomographic symbols were anticipated in the work
\cite{oman'ko-97} and elaborated in
\cite{vitale::dual,omanko:vitale}. Quantumness of qudit states was
demonstrated by means of dual tomographic symbols in
\cite{filipp-quant-witness}.

\section{\label{section:star:product:kernel} Kernel of star-product for spin tomographic symbols}

Since operators and tomographic symbols are in strong relation
with each other, any operation on $\hat{A}$ and $\hat{B}$
corresponds to an adequate operation on functions
$f_{\hat{A}}({\bf x})$ and $f_{\hat{B}}({\bf x})$. For instance,
the sum of $\hat{A}$ and $\hat{B}$ maps to the sum of
$f_{\hat{A}}({\bf x})$ and $f_{\hat{B}}({\bf x})$. As far as the
product $\hat{A}\hat{B}$ is concerned, the corresponding
tomographic symbol $f_{\hat{A}\hat{B}}({\bf x})$ is called the
star-product of the symbols $f_{\hat{A}}({\bf x})$ and
$f_{\hat{B}}({\bf x})$:

\begin{equation}
f_{\hat{A}\hat{B}}({\bf x}) = f_{\hat{A}}({\bf x}) \star
f_{\hat{B}}({\bf x}).
\end{equation}

\begin{figure}
\begin{center}
\includegraphics{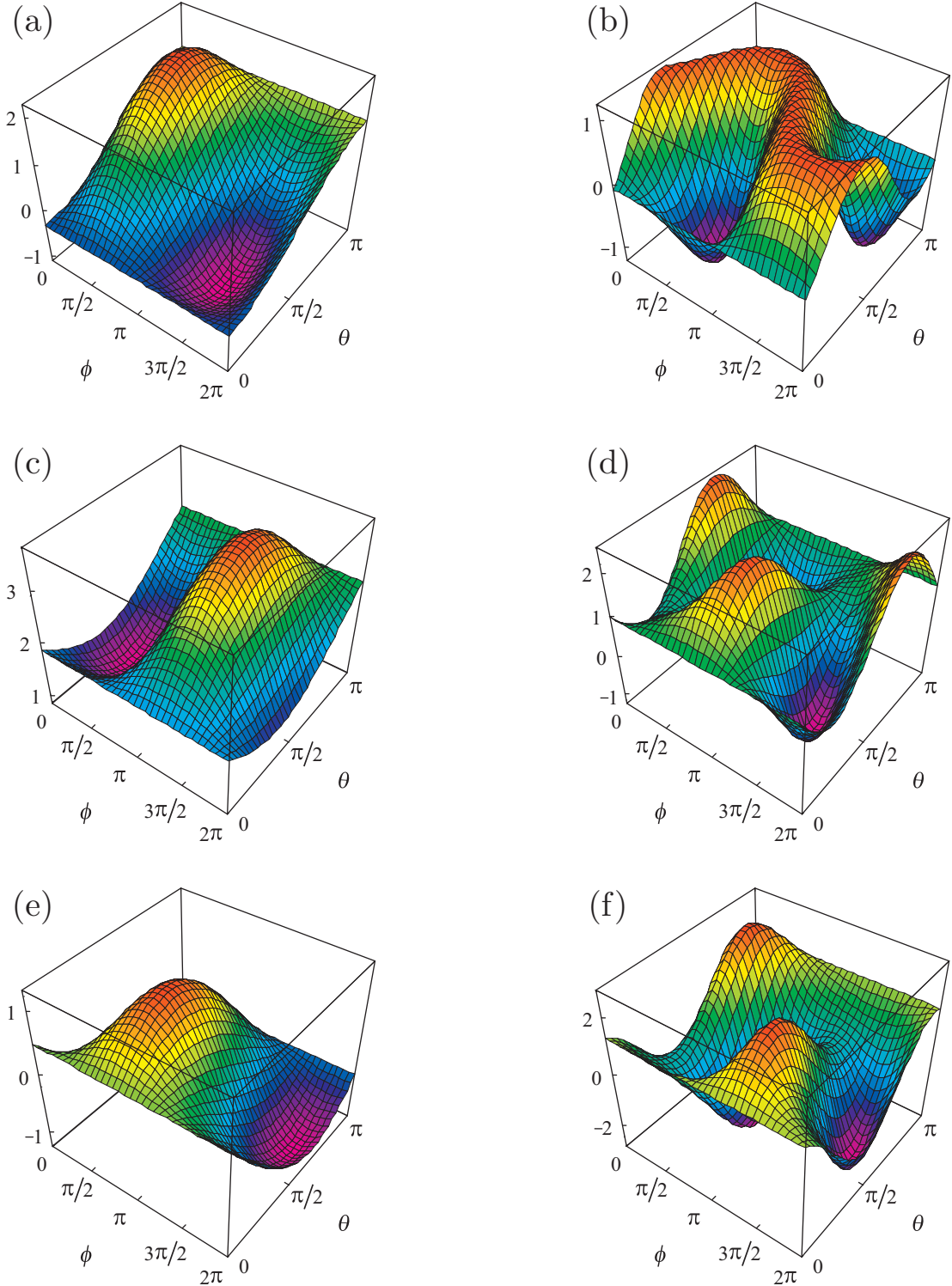}
\caption{\label{fig:combined} Non-locality of tomographic kernels.
Delta-function $K_{j}^{\delta}({\bf n}_2,{\bf
n}_1(\theta,\phi),m_2,m_1)$ on the set of qubit tomograms (a) and
on the set of qutrit tomograms (b). Star-product kernel
$K_{j}({\bf n}_3,{\bf n}_2,{\bf n}_1 (\theta,\phi),m_3,m_2,m_1)$
for qubits: real (c) and image (e) parts; and that for qutrits:
real (d) and image (f) parts. In (a) we set ${\bf n}_2 =
(0,-\sqrt{3}/2,1/2)$, $m_2=-1/2$, and $m_1=1/2$. In (b) ${\bf n}_2
= (-1/2\sqrt{2},\sqrt{3}/2\sqrt{2},1/\sqrt{2})$, $m_2=1$, and
$m_1=0$. In both (c) and (e) ${\bf n}_3 = (-1/2,-\sqrt{3}/2,0)$,
${\bf n}_2 =
(-\sqrt{3}/2\sqrt{2},-\sqrt{3}/2\sqrt{2},-1/\sqrt{2})$, and
$m_3=m_2=m_1=1/2$. Finally in both (d) and (f) we use ${\bf n}_3 =
(0,1,0)$, ${\bf n}_2 = (1/2,-1/2,1/\sqrt{2})$, $m_3=-1$, $m_2=1$,
and $m_1=0$.}
\end{center}
\end{figure}

\noindent By definition, one has

\begin{equation}
f_{\hat{A}\hat{B}}({\bf x}_{1}) = {\rm Tr} \left(
\hat{A}\hat{B}\hat{U}_{j}({\bf x}_{1}) \right) = \int\!\!\!\int
f_{\hat{A}}({\bf x}_{3}) f_{\hat{B}}({\bf x}_{2}) K_{j}({\bf
x}_{3},{\bf x}_{2},{\bf x}_{1}) d {\bf x}_{2} d {\bf x}_{3},
\end{equation}

\noindent where the function

\begin{equation}
\label{kernel:general} K_{j}({\bf x}_{3},{\bf x}_{2},{\bf x}_{1})
= {\rm Tr} \left( \hat{D}_{j}({\bf x}_{3})\hat{D}_{j}({\bf
x}_{2})\hat{U}_{j}({\bf x}_{1}) \right)
\end{equation}

\noindent is called the kernel of the star-product scheme. It is
worth noting that this kernel is non-local. The non-locality of
kernels of this type can also be illustrated if we consider
delta-function on the tomogram set. In fact, definitions
(\ref{tomogram}) and (\ref{rho:reconstruction}) are followed by
the relation

\begin{equation}
w_{j}({\bf x}_{1}) = \int w_{j}({\bf x}_{2}) {\rm Tr}\left(
\hat{D}_{j}({\bf x}_{2}) \hat{U}_{j}({\bf x}_{1}) \right) d {\bf
x}_{2},
\end{equation}

\noindent which holds true for an arbitrary spin tomogram
$w_{j}({\bf x})$. This implies that the function

\begin{equation}
K_{j}^{\delta} ({\bf x}_2, {\bf x}_1) = {\rm Tr}\left(
\hat{D}_{j}({\bf x}_{2}) \hat{U}_{j}({\bf x}_{1}) \right)
\end{equation}

\noindent can be treated as the kernel of the unity operator on
the set of spin tomograms and plays the role of an analogue of the
Dirac delta-function. Some examples of the star-product kernels
and the delta-functions for the low-spin states are depicted in
Fig. \ref{fig:combined}. It is readily seen that apart from being
non-local, the delta-function on the tomogram set is not
non-negative either.

%%%%%%%%%%%%%%%%%%%%%%%%%%%%%%%%%%%%%%%%%%%%%%%%%%%%%%%%%%%%%%%%%%%
%%%%%%%%%%%%%%%%%%%%%%%%%%%%%%%%%%%%%%%%%%%%%%%%%%%%%%%%%%%%%%%%%%%
%%%%%%%%%%%%%%%%%%%%%%%%%%%%%%%%%%%%%%%%%%%%%%%%%%%%%%%%%%%%%%%%%%%
%%%%%%%%%%%%%%%%%%%%%%%%%%%%%%%%%%%%%%%%%%%%%%%%%%%%%%%%%%%%%%%%%%%

By analogy with ordinary tomographic symbols, one can also
introduce the dual spin-tomographic star-product

\begin{equation}
f_{\hat{A}\hat{B}}^{d}({\bf x}) = f_{\hat{A}}^{d}({\bf x}) \star
f_{\hat{B}}^{d}({\bf x})
\end{equation}

\noindent with the non-local kernel of the form

\begin{equation}
\label{dual:kernel:general} K_{j}^{d}({\bf x}_{3},{\bf x}_{2},{\bf
x}_{1}) = {\rm Tr} \left( \hat{U}_{j}({\bf x}_{3})\hat{U}_{j}({\bf
x}_{2})\hat{D}_{j}({\bf x}_{1}) \right).
\end{equation}

Let us calculate the explicit form of the star-product kernel
(\ref{kernel:general}) for qudits with spin $j$.

Using the exponential representation of the dequantizer
(\ref{dequantizer:exponential}) and the quantizer
(\ref{quantizer:exponential}), we obtain

\begin{eqnarray}
\label{kernel:exponential:long} && K_{j}({\bf x}_{3},{\bf
x}_{2},{\bf x}_{1}) = \frac{(2j+1)^{2}}{(2\pi)^{3}}
\sum\limits_{s_{2}=-1}^{1} \sum\limits_{s_{3}=-1}^{1}
\frac{1}{(1-3s_{2}^{2})(1-3s_{3}^{2})}
\nonumber\\
&& \times \int\limits_{0}^{2\pi} \int\limits_{0}^{2\pi}
\int\limits_{0}^{2\pi} {\rm Tr} \left(e^{- i ({\bf n}_{3} \cdot
\hat{\bf J}) \varphi_{3}} e^{- i ({\bf n}_{2} \cdot \hat{\bf J})
\varphi_{2}} e^{- i ({\bf n}_{1} \cdot \hat{\bf J}) \varphi_{1}}
\right)  e^{i m_{1} \varphi_{1}} e^{i(m_{2}+s_{2})\varphi_{2}}
e^{i(m_{3}+s_{3})\varphi_{3}} d\varphi_{1} d\varphi_{2}
d\varphi_{3}.~~~
\end{eqnarray}

\noindent From this it follows that the spin-tomographic
star-product kernel (\ref{kernel:general}) is nothing else but the
Fourier transform of $SU(2)$ irreducible representation character

\begin{equation}
\label{character:trace} \chi ({\bf n}_{3},{\bf n}_{2},{\bf
n}_{1},\varphi_3,\varphi_2,\varphi_1) = {\rm Tr} \left(e^{- i
({\bf n}_{3} \cdot \hat{\bf J}) \varphi_{3}} e^{- i ({\bf n}_{2}
\cdot \hat{\bf J}) \varphi_{2}} e^{- i({\bf n}_{1} \cdot \hat{\bf
J}) \varphi_{1}} \right) = {\rm Tr} \left(e^{- i ({\mathfrak N}
\cdot \hat{\bf J}) \Phi} \right),
\end{equation}

\noindent where $\Phi = \Phi ({\bf n}_{1},{\bf n}_{2},{\bf n}_{3},
\varphi_{1}, \varphi_{2}, \varphi_{3})$ and ${\mathfrak N} =
{\mathfrak N} ({\bf n}_{1},{\bf n}_{2},{\bf n}_{3}, \varphi_{1},
\varphi_{2}, \varphi_{3})$ are respectively the angle and axis of
the resulting rotation which is equivalent to successive rotations
around axis ${\bf n}_{k}$ by angle $\varphi_{k}$, $k=1,2,3$.

To simplify formulas let us introduce the 3-vector ${\boldsymbol
\varphi}$ with components $(\varphi_{1},\varphi_{2},\varphi_{3})$
and the 9-vector ${\bf N}$ with components constructed from
components of three vectors ${\bf n}_{1}$, ${\bf n}_{2}$, ${\bf
n}_{3}$, i.e., ${\bf N} = ({\bf n}_{1}, {\bf n}_{2}, {\bf
n}_{3})$. Also, we designate

\begin{equation}
\int d {\boldsymbol \varphi} = \int\limits_{0}^{2\pi}
\int\limits_{0}^{2\pi} \int\limits_{0}^{2\pi} d\varphi_{1}
d\varphi_{2} d\varphi_{3}, \qquad {\bf m} = (m_1,m_2,m_3).
\end{equation}

Given the angle $\Phi$, the character has a rather simple form

\begin{equation}
\label{character:Chebyshev} \chi (\Phi) = \sum\limits_{m=-j}^{j}
e^{i m \Phi} = \frac{\sin ((2j+1) \Phi / 2)}{\sin (\Phi / 2)} =
U_{2j} \Big(\cos (\Phi/2)\Big),
\end{equation}

\noindent where $U_{n}(\cos
\theta)=\frac{\sin{(n+1)\theta}}{\sin\theta}$ is the Chebyshev
polynomial of the second kind of degree $n$
\cite{bateman,gradstein}.

Thus, combining
(\ref{kernel:exponential:long})--(\ref{character:Chebyshev}), we
obtain the integral representation of the star-product kernel

\begin{equation}
\label{kernel:Chebyshev} K_{j}({\bf x}_{3},{\bf x}_{2},{\bf
x}_{1}) = \frac{(2j+1)^{2}}{(2\pi)^{3}} \sum\limits_{s_{2}=-1}^{1}
\sum\limits_{s_{3}=-1}^{1} \frac{1}{(1-3s_{2}^{2})(1-3s_{3}^{2})}
I_{j}({\bf x}_{3},{\bf x}_{2},{\bf x}_{1}),
\end{equation}

\noindent where by $I_{j}({\bf x}_{3},{\bf x}_{2},{\bf x}_{1})$ we
denote the following integral:

\begin{equation}
\label{integral:in:kernel:Chebyshev} I_{j}({\bf x}_{3},{\bf
x}_{2},{\bf x}_{1}) = \int U_{2j} \left( \cos \frac{ \Phi ({\bf
N},{\boldsymbol \varphi})}{2} \right) e^{i ( {\bf m} \cdot
{\boldsymbol \varphi} ) } e^{i s_2 \varphi_2} e^{i s_3 \varphi_3}
d {\boldsymbol \varphi}.
\end{equation}

\noindent This implies that the kernel of spin-tomographic
star-product can be treated as the Fourier transform of the
Chebyshev polynomial of a specific argument. In the same way, it
can easily be checked that the kernel of the dual spin-tomographic
star-product reads

\begin{eqnarray}
\label{kernel:dual:Chebyshev} K_{j}^{d}({\bf x}_{3},{\bf
x}_{2},{\bf x}_{1}) = \frac{2j+1}{(2\pi)^{3}}
\sum\limits_{s_{1}=-1}^{1} \frac{1}{1-3s_{1}^{2}} \int U_{2j}
\left( \cos \frac{ \Phi ({\bf N},{\boldsymbol \varphi})}{2}
\right) e^{i ( {\bf m} \cdot {\boldsymbol \varphi} ) } e^{i s_1
\varphi_1} d {\boldsymbol \varphi}.
\end{eqnarray}

Further, the angle $\Phi$ does not depend on spin $j$.
Consequently it is possible to calculate it for qubits and then
extend the obtained result to other spins. Substituting $1/2$ for
$j$ in (\ref{character:Chebyshev}), we get the character for
qubits

\begin{equation}
\chi_{1/2} (\Phi) =  U_{1} \Big(\cos (\Phi/2)\Big) = 2 \cos
(\Phi/2).
\end{equation}

\noindent On the other hand, from (\ref{character:trace}) it
follows that

\begin{equation}
\chi_{1/2} (\Phi) =  {\rm Tr} \left(e^{- i({\bf n}_{3} \cdot
\hat{\boldsymbol\sigma}) \varphi_{3}/2} e^{- i({\bf n}_{2} \cdot
\hat{\boldsymbol\sigma}) \varphi_{2}/2} e^{- i({\bf n}_{1} \cdot
\hat{\boldsymbol\sigma}) \varphi_{1}/2} \right),
\end{equation}

\noindent where $\hat{\boldsymbol\sigma} =
(\hat{\sigma}_{x},\hat{\sigma}_{y},\hat{\sigma}_{z})$ is a set of
the Pauli matrices. Employing the known property of Pauli matrices
$\hat{\sigma}_{\alpha} \hat{\sigma}_{\beta} = \delta_{\alpha\beta}
\hat{1} + i \varepsilon_{\alpha\beta\gamma}
\hat{\sigma}_{\gamma}$, it is not hard to prove that the relations

\begin{eqnarray}
& ({\bf a} \cdot \hat{\boldsymbol\sigma}) ({\bf b} \cdot
\hat{\boldsymbol\sigma}) = a_{\alpha} b_{\beta}
\hat{\sigma}_{\alpha} \hat{\sigma}_{\beta} = ({\bf a} \cdot {\bf
b}) \hat{1} + i \Big( [{\bf a} \times {\bf b}] \cdot
\hat{\boldsymbol\sigma} \Big), \\
& \label{exp:sigma} e^{-i({\bf n} \cdot \hat{\boldsymbol\sigma})
\varphi /2} = \hat{1} \cos (\varphi / 2) -i ({\bf n} \cdot
\hat{\boldsymbol\sigma}) \sin (\varphi /2)
\end{eqnarray}

\noindent are valid whenever ${\bf n}^{2}$ is equal to unity. In
view of these relations, we finally obtain

\begin{equation}
e^{- i({\bf n}_{3} \cdot \hat{\boldsymbol\sigma}) \varphi_{3}/2}
e^{- i({\bf n}_{2} \cdot \hat{\boldsymbol\sigma}) \varphi_{2}/2}
e^{- i({\bf n}_{1} \cdot \hat{\boldsymbol\sigma}) \varphi_{1}/2} =
\hat{1} \cos \left( \Phi ({\bf N}, {\boldsymbol \varphi}) / 2
\right) -i ( {\mathfrak{N}} \cdot \hat{\boldsymbol\sigma}) \sin
\left( \Phi ({\bf N}, {\boldsymbol \varphi}) / 2 \right).
\end{equation}

\noindent Recall that the angle $\Phi$ depends on three rotation
angles $\varphi_{1}$, $\varphi_{2}$, $\varphi_{3}$ and three
directions ${\bf n}_{1}$, ${\bf n}_{2}$, ${\bf n}_{3}$. Departing
from this notation, one can easily derive the resulting rotation
angle

\begin{eqnarray}
&& \cos \left( \Phi ({\bf N}, {\boldsymbol \varphi}) / 2 \right) =
\cos ({\varphi_{1}}/{2}) \cos ({\varphi_{2}}/{2}) \cos
({\varphi_{3}}/{2}) -({\bf n}_{1} \cdot {\bf n}_{2}) \sin
({\varphi_{1}}/{2}) \sin ({\varphi_{2}}/{2}) \cos ({\varphi_{3}}/{2}) \nonumber\\
&& -({\bf n}_{2} \cdot {\bf n}_{3}) \cos ({\varphi_{1}}/{2}) \sin
({\varphi_{2}}/{2}) \sin ({\varphi_{3}}/{2}) -({\bf n}_{3} \cdot
{\bf n}_{1}) \sin ({\varphi_{1}}/{2}) \cos
({\varphi_{2}}/{2}) \sin ({\varphi_{3}}/{2}) \nonumber\\
&& + \left({\bf n}_{1} \cdot \left[ {\bf n}_{2} \times {\bf n}_{3}
\right] \right) \sin ({\varphi_{1}}/{2}) \sin ({\varphi_{2}}/{2})
\sin ({\varphi_{3}}/{2})
\end{eqnarray}

\noindent and the rotation axis

\begin{eqnarray} && \mathfrak{N} \sin \left( \Phi
({\bf N}, {\boldsymbol \varphi}) / 2 \right) = {\bf n}_{1} \sin
({\varphi_{1}}/{2}) \cos
({\varphi_{2}}/{2}) \cos ({\varphi_{3}}/{2}) \nonumber\\
&& + {\bf n}_{2} \cos ({\varphi_{1}}/{2}) \sin ({\varphi_{2}}/{2})
\cos ({\varphi_{3}}/{2}) + {\bf n}_{3} \cos ({\varphi_{1}}/{2})
\cos ({\varphi_{2}}/{2}) \sin ({\varphi_{3}}/{2}) \nonumber\\
&& - \Big\{ {\bf n}_{1} ({\bf n}_{2} \cdot {\bf n}_{3}) - {\bf
n}_{2} ({\bf n}_{1} \cdot {\bf n}_{3}) + {\bf n}_{3} ({\bf n}_{1}
\cdot {\bf n}_{2})\Big\} \sin ({\varphi_{1}}/{2}) \sin
({\varphi_{2}}/{2}) \sin ({\varphi_{3}}/{2}) \nonumber\\
&& - [{\bf n}_{1} \times {\bf n}_{2}] \sin ({\varphi_{1}}/{2})
\sin ({\varphi_{2}}/{2}) \cos ({\varphi_{3}}/{2}) - [{\bf n}_{2}
\times {\bf n}_{3}] \cos ({\varphi_{1}}/{2}) \sin
({\varphi_{2}}/{2}) \sin ({\varphi_{3}}/{2}) \nonumber\\
&& - [{\bf n}_{1} \times {\bf n}_{3}] \sin ({\varphi_{1}}/{2})
\cos ({\varphi_{2}}/{2}) \sin ({\varphi_{3}}/{2}).
\end{eqnarray}

\begin{figure}
\begin{center}
\includegraphics{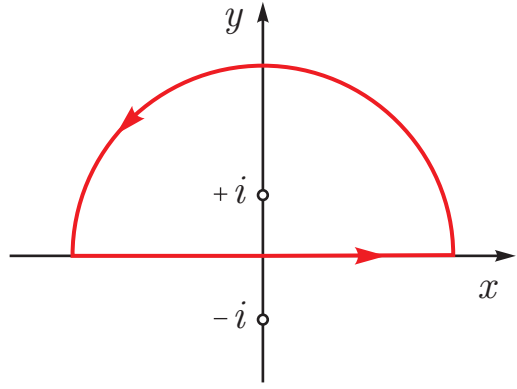}
\caption{\label{fig:contour} Path of integration used for
computation of the integral (\ref{I:intermediate}).}
\end{center}
\end{figure}

Now, when $\cos \left( \Phi ({\bf N}, {\boldsymbol \varphi}) / 2
\right)$ is known, integral (\ref{integral:in:kernel:Chebyshev})
can be evaluated. Change of variables $t_k = -\cot(\varphi_k/2)$,
$k=1,2,3$ results in the integral taking the form

\begin{eqnarray}
\label{I:intermediate} && I_{j}({\bf x}_{3},{\bf x}_{2},{\bf
x}_{1}) = 8 \int \!\!\!\!\! \int\limits_{-\infty}^{+\infty}
\!\!\!\!\! \int \frac{d t_1 d t_2 d t_3 (t_3 - i)^{m_3+s_3-1} (t_2
- i)^{m_2+s_2-1} (t_1 - i)^{m_1-1} }{ (t_3 + i)^{m_3+s_3+1} (t_2 +
i)^{m_2+s_2+1} (t_1 + i)^{m_1+1} } \nonumber\\
&& \qquad \times U_{2j} \left( \frac{t_1 t_2 t_3 - t_1 ({\bf
n}_{2} \cdot {\bf n}_{3}) - t_2 ({\bf n}_{3} \cdot {\bf n}_{1}) -
t_3 ({\bf n}_{1} \cdot {\bf n}_{2}) - \left({\bf n}_{1} \cdot
\left[ {\bf n}_{2} \times {\bf n}_{3} \right] \right) }{
(t_{1}-i)^{1/2}(t_{1}+i)^{1/2}(t_{2}-i)^{1/2}(t_{2}+i)^{1/2}(t_{3}-i)^{1/2}(t_{3}+i)^{1/2}}
\right),
\end{eqnarray}

\noindent where $z^{1/2}$, $z \in \mathbb{C}$ is regarded as a
principal branch of the square root function, with the branch cut
being along the positive real axis. Since the integrand decreases
fast enough as $|t_k| \rightarrow \infty$, $k=1,2,3$, one can
calculate the integral in question with the help of the residue
theorem. Indeed, choosing for each complex variable $t_k$,
$k=1,2,3$ the path of integration shown in Fig. \ref{fig:contour},
we have $I_{j}({\bf x}_{3},{\bf x}_{2},{\bf x}_{1})=(2\pi i)^{3}
{\rm Res}_{t_1 = i} {\rm Res}_{t_2 = i} {\rm Res}_{t_3 = i}$. In
order to calculate the residues one needs to find a coefficient
corresponding to the term $(t_1-i)^{-1}(t_2-i)^{-1}(t_3-i)^{-1}$.

%%%%%%%%%%%%%%%%%%%%%%%%%%%%%%%%%%%%%%%%%%%%%%%%%%%%%%%%%%%%%%%%%%%
%%%%%%%%%%%%%%%%%%%%%%%%%%%%%%%%%%%%%%%%%%%%%%%%%%%%%%%%%%%%%%%%%%%
%%%%%%%%%%%%%%%%%%%%%%%%%%%%%%%%%%%%%%%%%%%%%%%%%%%%%%%%%%%%%%%%%%%
%%%%%%%%%%%%%%%%%%%%%%%%%%%%%%%%%%%%%%%%%%%%%%%%%%%%%%%%%%%%%%%%%%%

Employing the expansion of the Chebyshev polynomial

\begin{equation}
U_{2j}(x) = \sum\limits_{k=0}^{[j]}
\frac{(-1)^{k}(2j-k)!}{k!(2j-2k)!} (2x)^{2j-2k},
\end{equation}

\noindent the integral (\ref{I:intermediate}) can be written in
the form

\begin{eqnarray}
\label{integral:residues} && I_{j}({\bf x}_{3},{\bf x}_{2},{\bf
x}_{1}) = \sum\limits_{k=0}^{[j]} \frac{(-1)^{k}(2j-k)!
2^{2j-2k+3} }{ k!(2j-2k)!} \int \!\!\!\!\!
\int\limits_{-\infty}^{+\infty}
\!\!\!\!\! \int d t_1 d t_2 d t_3 f^{2j-2k}(t_1,t_2,t_3) \nonumber\\
&& \qquad\qquad\qquad\qquad \times \frac{ (t_3 +
i)^{-j-m_3-s_3+k-1} (t_2 + i)^{-j-m_2-s_2+k-1} (t_1 +
i)^{-j-m_1+k-1}
 }{ (t_3 - i)^{j-m_3-s_3-k+1} (t_2 -
i)^{j-m_2-s_2-k+1} (t_1 - i)^{j-m_1-k+1} } ,
\end{eqnarray}

\noindent where the $(2j-2k)$th power of the function

\begin{eqnarray}
\label{f} f(t_1,t_2,t_3) = && (t_1-i)(t_2-i)(t_3-i)
+i(t_1-i)(t_2-i)+i(t_2-i)(t_3-i)+i(t_3-i)(t_1-i) \nonumber\\
&& - (t_1-i) \Big(1+({\bf n}_{2} \cdot {\bf n}_{3}) \Big) -
(t_2-i) \Big( 1+({\bf n}_{3} \cdot {\bf n}_{1}) \Big) - (t_3-i)
\Big( 1+({\bf n}_{1} \cdot {\bf
n}_{2}) \Big) \nonumber\\
&& - \left({\bf n}_{1} \cdot \left[ {\bf n}_{2} \times {\bf n}_{3}
\right] \right) - i \Big( 1+({\bf n}_{1} \cdot {\bf n}_{2})+({\bf
n}_{2} \cdot {\bf n}_{3})+({\bf n}_{3} \cdot {\bf n}_{1}) \Big)
\end{eqnarray}

\noindent resolves to the following sum:

\begin{eqnarray}
\label{f:expansion} && f^{2j-2k}(t_1,t_2,t_3) \nonumber\\ && =
\sum_{p_1+\dots+p_8=2j-2k}
\frac{(2j-2k)!}{\prod\nolimits_{l=1}^{8}p_{l}!}
(t_1-i)^{p_1+p_2+p_4+p_5}(t_2-i)^{p_1+p_2+p_3+p_6}(t_3-i)^{p_1+p_3+p_4+p_7} \nonumber\\
&& \times i^{p_2+p_3+p_4+2(p_5+p_6+p_7)+3p_8} \Big( 1+({\bf n}_{2}
\cdot {\bf n}_{3}) \Big)^{p_5} \Big( 1+({\bf n}_{3} \cdot {\bf
n}_{1}) \Big)^{p_6} \Big( 1+({\bf
n}_{1} \cdot {\bf n}_{2}) \Big)^{p_7} \nonumber\\
&& \times \Big( 1+({\bf n}_{1} \cdot {\bf n}_{2})+({\bf n}_{2}
\cdot {\bf n}_{3})+({\bf n}_{3} \cdot {\bf n}_{1}) - i \left({\bf
n}_{1} \cdot \left[ {\bf n}_{2} \times {\bf n}_{3} \right] \right)
\Big)^{p_8}.
\end{eqnarray}

Moreover, one should extract terms $(t_k - i)$ from the general
expression $(t_k + i)^r$. It is obvious that the expansion of
$(t+i)^{r}$ to the powers of $(t-i)$ reads

\begin{equation}
\label{t+i:expansion} (t+i)^{r} = (2i)^{r} \left( 1+\frac{t-i}{2i}
\right)^{r} = \sum\limits_{q=0}^{\infty} \left(%
\begin{array}{c}
  r \\
  q \\
\end{array}%
\right) \frac{(t-i)^{q}}{(2i)^{q-r}},
\end{equation}

\noindent where we introduced binomial coefficients according to
the rule \cite{korn-binomial}

\begin{equation}
\left(%
\begin{array}{c}
  r \\
  q \\
\end{array}%
\right) = \frac{r (r-1) \dots (r-q+1)}{q!} .
\end{equation}

\noindent Here $r$ is supposed to be real, $q$ is an integer, $\left(%
\begin{array}{c}
  r \\
  q \\
\end{array}%
\right) = 0$ if $q<0$, and $\left(%
\begin{array}{c}
  r \\
  0 \\
\end{array}%
\right) = 1$.

If we combine (\ref{integral:residues}) with (\ref{f:expansion})
and (\ref{t+i:expansion}), we can calculate the residues involved
and then the integral $I$. The direct computation yields

\begin{eqnarray}
\label{integral:residues:computed} && I_{j}({\bf x}_{3},{\bf
x}_{2},{\bf x}_{1}) \nonumber\\
&& = (2\pi)^{3} \sum\limits_{k=0}^{[j]} \sum_{p_1+\dots+p_8=2j-2k}
\frac{(-1)^{k}(2j-k)! 2^{2j-2k+3} }{ k! \prod_{l=1}^{8}p_{l}!}
\Big(1+({\bf n}_{2} \cdot {\bf n}_{3}) \Big)^{p_5} \Big( 1+({\bf
n}_{3} \cdot
{\bf n}_{1}) \Big)^{p_6} \nonumber\\
&& \times \Big( 1+({\bf n}_{1} \cdot {\bf n}_{2}) \Big)^{p_7}
\Big( 1+({\bf n}_{1} \cdot {\bf n}_{2})+({\bf n}_{2} \cdot {\bf
n}_{3})+({\bf n}_{3} \cdot {\bf n}_{1}) - i \left({\bf n}_{1}
\cdot \left[ {\bf
n}_{2} \times {\bf n}_{3} \right] \right) \Big)^{p_8} \nonumber\\
&& \times \sum\limits_{q_3=0}^{\infty}
\sum\limits_{q_2=0}^{\infty} \sum\limits_{q_1=0}^{\infty} \frac{
i^{p_2+p_3+p_4+2(p_5+p_6+p_7)+3p_8+3} }{
(2i)^{3j+m_3+m_2+m_1+s_3+s_2-3k+3+q_3+q_2+q_1} } \left(%
\begin{array}{c}
  -j-m_1+k-1 \\
  q_1 \\
\end{array}%
\right) \nonumber\\
&& \times \left(%
\begin{array}{c}
  -j-m_2-s_2+k-1 \\
  q_2 \\
\end{array}%
\right) \left(%
\begin{array}{c}
  -j-m_3-s_3+k-1 \\
  q_3 \\
\end{array}%
\right)  \nonumber\\
&& \times \delta_{j-m_1-k-p_1-p_2-p_4-p_5-q_1,0}
\delta_{j-m_2-s_2-k-p_1-p_2-p_3-p_6-q_2,0}
\delta_{j-m_3-s_3-k-p_1-p_3-p_4-p_7-q_3,0}.
\end{eqnarray}

\noindent Finally, substituting the calculated value of $I$ for
the integral in (\ref{kernel:Chebyshev}), we obtain the explicit
form of the spin-tomographic star-product kernel

\begin{eqnarray}
\label{kernel:final:explicit:expression} && K_{j}({\bf x}_{3},{\bf x}_{2},{\bf x}_{1}) \nonumber\\
&& = (2j+1)^{2} \sum\limits_{s_{3}=-1}^{1}
\sum\limits_{s_{2}=-1}^{1} \sum\limits_{k=0}^{[j]}
\sum\limits_{p_{1}+\dots+p_{8}=2j-2k} \frac{(-1)^{k} (2j-k)!
}{(1-3s_{3}^{2})(1-3s_{2}^{2})
2^{-p_1+p_5+p_6+p_7+2p_8} k! \prod_{l=1}^{8}p_{l}! } \nonumber\\
&& \qquad \times \Big(1+({\bf n}_{2} \cdot {\bf n}_{3})\Big)^{p_5}
\Big(1+({\bf n}_{3} \cdot {\bf n}_{1})\Big)^{p_6} \Big(1+({\bf
n}_{1} \cdot {\bf n}_{2})\Big)^{p_7} \nonumber\\
&& \qquad \times \Big(1 +({\bf n}_{1} \cdot {\bf n}_{2}) +({\bf
n}_{2} \cdot {\bf n}_{3}) +({\bf n}_{3} \cdot {\bf n}_{1}) - i
\left({\bf n}_{1} \cdot \left[ {\bf n}_{2} \times {\bf n}_{3}
\right] \right)
\Big)^{p_8} \nonumber\\
&& \qquad \times \left(%
\begin{array}{c}
  -j-m_1+k-1 \\
  j-m_1-k-p_1-p_2-p_4-p_5 \\
\end{array}%
\right) \left(%
\begin{array}{c}
  -j-m_2-s_2+k-1 \\
  j-m_2-s_2-k-p_1-p_2-p_3-p_6 \\
\end{array}%
\right) \nonumber\\
&& \qquad \times \left(%
\begin{array}{c}
  -j-m_3-s_3+k-1 \\
  j-m_3-s_3-k-p_1-p_3-p_4-p_7 \\
\end{array}%
\right).
\end{eqnarray}

\section{\label{section:equivalency}Equivalency of kernel representations}

The problem of the explicit form of the star-product kernel has
been attacked from different perspectives. In the work
\cite{castanos}, the explicit formula is expressed in terms of
Clebsch-Gordan and Racah coefficients. To be more precise, the
result is

\begin{eqnarray}
\label{kernel:eplicit:C-G} && \mathcal{K}_{j}'({\bf x}_{3},{\bf
x}_{2},{\bf x}_{1}) = (-1)^{j-m_1-m_2-m_3}
\sum\limits_{L_1=0}^{2j}
\sum\limits_{L_2=0}^{2j} \sum\limits_{L_3=0}^{2j} (-1)^{L_1+L_2+L_3} \sqrt{(2L_3+1)^{3}(2L_2+1)^{3}(2L_1+1)} \nonumber\\
&& \times \langle jm_1 ; j -m_1 | L_1 0 \rangle \langle jm_2 ; j -m_2 | L_2 0 \rangle \langle jm_3 ; j -m_3 | L_3 0 \rangle \left\{%
\begin{array}{ccc}
  L_2 & L_3 & L_1 \\
  j & j & j \\
\end{array}%
\right\} \nonumber\\
&& \times \sum\limits_{M_1 = -L_1}^{L_1} \sum\limits_{M_2 = -L_2}^{L_2} \sum\limits_{M_3 = -L_3}^{L_3} \left(%
\begin{array}{ccc}
  L_2 & L_3 & L_1 \\
  M_2 & M_3 & M_1 \\
\end{array}%
\right) \mathcal{D}_{0 -M_1}^{(L_1)}(0,\theta_1,-\phi_1)
\mathcal{D}_{0 -M_2}^{(L_2)}(0,\theta_2,-\phi_2) \mathcal{D}_{0
-M_3}^{(L_3)}(0,\theta_3,-\phi_3),\nonumber\\
\end{eqnarray}

\noindent where the Wigner $\mathcal{D}$-function reads

\begin{eqnarray}
\label{D-Wigner:function} \mathcal{D}_{m' m}^{(j)}
(\alpha,\beta,\gamma) = e^{-im' \alpha} e^{-im \gamma}
\sum\limits_{s} &&
\frac{(-1)^{s}\sqrt{(j+m)!(j-m)!(j+m')!(j-m')!}}
{s!(j-m'-s)!(j+m-s)!(m'-m+s)!} \nonumber \\ && \times
\left(\cos\frac{\beta}{2}\right)^{2j+m-m'-2s}
\left(-\sin\frac{\beta}{2}\right)^{m'-m+2s}.
\end{eqnarray}

\noindent In \cite{filipp-spin-tomography}, using the irreducible
tensor operators for the $SU(2)$ group \cite{varshalovich,klimov},
the same result is specified for the low-spin states and presented
in the form of the expansion to orthogonal summands. In the
present work, starting from the exponential representation of the
quantizer and dequantizer operators, we managed to obtain another
explicit form of the star-product kernel which can also be
presented in the form

\begin{eqnarray}
\label{kernel:explicit:new} && \mathcal{K}_{j}''({\bf x}_{3},{\bf
x}_{2},{\bf x}_{1})
\nonumber\\
&& = (2j+1)^{2} \sum\limits_{s_{3}=-1}^{1}
\sum\limits_{s_{2}=-1}^{1} \sum\limits_{k=0}^{[j]}
\sum\limits_{p_{1}+\dots+p_{8}=2j-2k} \frac{(-1)^{k} (2j-k)!
}{(1-3s_{3}^{2})(1-3s_{2}^{2})
2^{-p_1+p_5+p_6+p_7+2p_8} k! \prod_{l=1}^{8}p_{l}! } \nonumber\\
&& \times \Big(1 + ( {\bf n}(\theta_2,\phi_2) \cdot {\bf
n}(\theta_3,\phi_3) ) \Big)^{p_5} \Big(1+({\bf n}(\theta_3,\phi_3)
\cdot {\bf n}(\theta_1,\phi_1) )\Big)^{p_6} \Big(1+({\bf
n}(\theta_1,\phi_1) \cdot
{\bf n}(\theta_2,\phi_2) )\Big)^{p_7} \nonumber\\
&& \times \Big(1 +({\bf n}(\theta_1,\phi_1) \cdot {\bf
n}(\theta_2,\phi_2)) +({\bf n}(\theta_2,\phi_2) \cdot {\bf
n}(\theta_3,\phi_3)) +({\bf n}(\theta_3,\phi_3) \cdot {\bf
n}(\theta_1,\phi_1)) \nonumber\\
&& \qquad - i \left({\bf n}(\theta_1,\phi_1) \cdot \left[ {\bf
n}(\theta_2,\phi_2) \times {\bf n}(\theta_3,\phi_3) \right]
\right)
\Big)^{p_8} \nonumber\\
&& \times \left(%
\begin{array}{c}
  -j-m_1+k-1 \\
  j-m_1-k-p_1-p_2-p_4-p_5 \\
\end{array}%
\right) \left(%
\begin{array}{c}
  -j-m_2-s_2+k-1 \\
  j-m_2-s_2-k-p_1-p_2-p_3-p_6 \\
\end{array}%
\right) \nonumber\\
&& \times \left(%
\begin{array}{c}
  -j-m_3-s_3+k-1 \\
  j-m_3-s_3-k-p_1-p_3-p_4-p_7 \\
\end{array}%
\right).
\end{eqnarray}

It is obvious that all the different formulas must be equivalent
on the set of tomograms. This is followed by a specific sum rule
for Clebsch-Gordan and Racah coefficients.  Namely

\begin{equation}
\mathcal{K}_{j}'({\bf x}_{3},{\bf x}_{2},{\bf x}_{1}) \sim
\mathcal{K}_{j}''({\bf x}_{3},{\bf x}_{2},{\bf x}_{1}),
\end{equation}

\noindent where the sign $\sim$ is defined through a biconditional
implication of the form

\begin{eqnarray}
&& \mathcal{K}_{j}'({\bf x}_{3},{\bf x}_{2},{\bf x}_{1}) \sim
\mathcal{K}_{j}''({\bf x}_{3},{\bf x}_{2},{\bf x}_{1})
\Longleftrightarrow  \Bigg\{ \int\!\!\!\int f_{\hat{A}}({\bf
x}_{3}) f_{\hat{B}}({\bf x}_{2}) \mathcal{K}_{j}'({\bf x}_{3},{\bf
x}_{2},{\bf x}_{1}) d {\bf x}_{2} d {\bf x}_{3} \nonumber\\
&& = \int\!\!\!\int f_{\hat{A}}({\bf x}_{3}) f_{\hat{B}}({\bf
x}_{2}) \mathcal{K}_{j}''({\bf x}_{3},{\bf x}_{2},{\bf x}_{1}) d
{\bf x}_{2} d {\bf x}_{3} {\rm ~ for ~ all ~ symbols ~ }
f_{\hat{A}}({\bf x}) {\rm ~ and ~ } f_{\hat{B}}({\bf x}) \Bigg\}.
\end{eqnarray}

\noindent Some of sum rules for Clebsch-Gordan coefficients can be
found in \cite{varshalovich,shelepin-fian,smorodinsky-shelepin}.

Though there takes place an ambiguity in the star-product kernel,
all types of the kernel must be equivalent for calculating the
symbol of the product of two given operators. As far as functions
(\ref{kernel:eplicit:C-G}) and (\ref{kernel:explicit:new}) are
concerned, in case of qubits ($j=1/2$), both formulas turned out
to be the same (and consequently equal to that found in
\cite{filipp-spin-tomography}). In case of qutrits ($j=1$), the
kernel (\ref{kernel:explicit:new}) contains more terms than the
kernel (\ref{kernel:eplicit:C-G}) expressed in terms of
Clebsch-Gordan coefficients. Actually, all redundant terms give
zero while integrating with tomographic symbols. In Appendix 1, we
discuss the cause of the deviation between kernels and present the
difference $\Delta_{j} = \mathcal{K}_{j}''({\bf x}_{3},{\bf
x}_{2},{\bf x}_{1}) - \mathcal{K}_{j}'({\bf x}_{3},{\bf
x}_{2},{\bf x}_{1})$ for qutrits ($j=1$).

\section{\label{section:recurrence}Recurrence relation for spin-tomographic kernels}

The Chebyshev polynomials obey the recurrence relation of the form
\cite{bateman,gradstein}

\begin{equation}
U_{n+1}(x) = 2x U_{n}(x) - U_{n-1}(x).
\end{equation}

\noindent Using this peculiar property of the Chebyshev
polynomials, it is easy to prove that there exists a similar
recurrence relation for integral
(\ref{integral:in:kernel:Chebyshev}). In fact, one has

\begin{equation}
I_{j+1/2}({\bf x}_3,{\bf x}_2,{\bf x}_1) = 2J_{j}({\bf x}_3,{\bf
x}_2,{\bf x}_1) - I_{j-1/2}({\bf x}_3,{\bf x}_2,{\bf x}_1),
\end{equation}

\noindent where

\begin{eqnarray}
&& J_{j}({\bf x}_3,{\bf x}_2,{\bf x}_1) = \sum\limits_{k=0}^{[j]}
\frac{(-1)^{k}(2j-k)! 2^{2j-2k+3} }{ k!(2j-2k)!} \int \!\!\!\!\!
\int\limits_{-\infty}^{+\infty}
\!\!\!\!\! \int d t_1 d t_2 d t_3 f^{2j-2k+1}(t_1,t_2,t_3) \nonumber\\
&& \qquad\qquad\qquad\qquad \times \frac{ (t_3 +
i)^{-j-m_3-s_3+k-3/2} (t_2 + i)^{-j-m_2-s_2+k-3/2} (t_1 +
i)^{-j-m_1+k-3/2}
 }{ (t_3 - i)^{j-m_3-s_3-k+3/2} (t_2 -
i)^{j-m_2-s_2-k+3/2} (t_1 - i)^{j-m_1-k+3/2} }.~~~~~
\end{eqnarray}

Employing the explicit form (\ref{f}) of the function
$f(t_1,t_2,t_3)$, one can calculate the integral involved just in
the same way as it was fulfilled before and then utilize the
following property of binomial coefficients \cite{korn-binomial}:

\begin{equation}
\left(%
\begin{array}{c}
  r \\
  q+1 \\
\end{array}%
\right) = \left(%
\begin{array}{c}
  r+1 \\
  q+1 \\
\end{array}%
\right) - \left(%
\begin{array}{c}
  r \\
  q \\
\end{array}%
\right).
\end{equation}

\noindent The result is

\begin{eqnarray}
&& J_{j}({\bf x}_3,{\bf x}_2,{\bf x}_1) =
\sum\limits_{m_1',m_2',m_3'} I_{j} ({\bf
N}, {\bf m}') \Bigg[ \delta_{m_1',m_1+1/2} ~ \delta_{m_2',m_2+1/2} ~ \delta_{m_3',m_3+1/2} \nonumber\\
&& +\frac{1}{2} \sum\limits_{k<l} \sum\limits_{h \neq k,l}
\sum\limits_{\nu = -1/2}^{1/2} (-1)^{1/2+\nu}
\delta_{m_h',m_h+\nu} ~ \delta_{m_k',m_k+1/2} ~
\delta_{m_l',m_l+1/2} \nonumber\\
&& +\frac{1}{4} \sum\limits_{k<l} \sum\limits_{h \neq k,l} \Big( 1
+ ({\bf n}_k \cdot {\bf n}_l) \Big) \sum\limits_{\nu_k =
-1/2}^{1/2} \sum\limits_{\nu_l = -1/2}^{1/2}
(-1)^{1+\nu_k+\nu_l} \delta_{m_h',m_h+1/2} ~ \delta_{m_k',m_k+\nu_k} ~ \delta_{m_l',m_l+\nu_l} \nonumber\\
&& +\frac{1}{8} \Big( 1 +({\bf n}_{1} \cdot {\bf n}_{2}) +({\bf
n}_{2} \cdot {\bf n}_{3}) +({\bf n}_{3} \cdot {\bf n}_{1}) - i
\left({\bf n}_{1} \cdot \left[ {\bf n}_{2} \times {\bf n}_{3}
\right] \right) \Big) \nonumber\\
&& \qquad \times \sum\limits_{\nu_1 = -1/2}^{1/2}
\sum\limits_{\nu_2 = -1/2}^{1/2} \sum\limits_{\nu_3 = -1/2}^{1/2}
(-1)^{3/2+\nu_1+\nu_2+\nu_3} \delta_{m_1',m_1+\nu_1} ~
\delta_{m_2',m_2+\nu_2} ~ \delta_{m_3',m_3+\nu_3} \Bigg].
\end{eqnarray}

Now if we recall (\ref{kernel:Chebyshev}), we obtain the
recurrence relation for spin-tomographic kernels. To be more
precise, the kernel for spin $(j+1/2)$ is expressed in terms of
kernels for spin $j$ and spin $(j-1/2)$ as follows:

\begin{eqnarray}
&& K_{j+1/2}({\bf x}_3,{\bf x}_2,{\bf x}_1) \equiv K_{j+1/2} ({\bf
N}, {\bf m}) %\nonumber\\ &&
= 2 \left( \frac{2j+2}{2j+1}
\right)^{2} \sum\limits_{m_1',m_2',m_3'} K_{j} ({\bf
N}, {\bf m}') \nonumber\\
&& \times \Bigg[ \delta_{m_1',m_1+1/2} ~ \delta_{m_2',m_2+1/2} ~
\delta_{m_3',m_3+1/2} %\nonumber\\ &&
+\frac{1}{2}
\sum\limits_{k<l} \sum\limits_{h \neq k,l} \sum\limits_{\nu =
-1/2}^{1/2} (-1)^{1/2+\nu} \delta_{m_h',m_h+\nu} ~
\delta_{m_k',m_k+1/2} ~
\delta_{m_l',m_l+1/2} \nonumber\\
&& +\frac{1}{4} \sum\limits_{k<l} \sum\limits_{h \neq k,l} \Big( 1
+ ({\bf n}_k \cdot {\bf n}_l) \Big) \sum\limits_{\nu_k =
-1/2}^{1/2} \sum\limits_{\nu_l = -1/2}^{1/2}
(-1)^{1+\nu_k+\nu_l} \delta_{m_h',m_h+1/2} ~ \delta_{m_k',m_k+\nu_k} ~ \delta_{m_l',m_l+\nu_l} \nonumber\\
&& +\frac{1}{8} \Big( 1 +({\bf n}_{1} \cdot {\bf n}_{2}) +({\bf
n}_{2} \cdot {\bf n}_{3}) +({\bf n}_{3} \cdot {\bf n}_{1}) - i
\left({\bf n}_{1} \cdot \left[ {\bf n}_{2} \times {\bf n}_{3}
\right] \right) \Big) \nonumber\\
&& \qquad \times \sum\limits_{\nu_1 = -1/2}^{1/2}
\sum\limits_{\nu_2 = -1/2}^{1/2} \sum\limits_{\nu_3 = -1/2}^{1/2}
(-1)^{3/2+\nu_1+\nu_2+\nu_3} \delta_{m_1',m_1+\nu_1} ~ \delta_{m_2',m_2+\nu_2} ~ \delta_{m_3',m_3+\nu_3} \Bigg] \nonumber\\
&& - \left( \frac{2j+2}{2j} \right)^{2} K_{j-1/2} ({\bf N}, {\bf
m}).
\end{eqnarray}

This recurrence formula reveals a special feature of the
star-product kernels. Indeed, the star-product kernel for qudits
with an arbitrary spin $j$ can be expressed in terms of the kernel
for qubits and that for spins equal to zero.

\section{\label{section:conclusions}Conclusions}

To resume we point out the main results of our work. We obtained
the explicit form of the star-product kernel for spin tomograms in
terms of Fourier transform of the Chebyshev polynomial (see Eqs.
(\ref{kernel:Chebyshev}) and
(\ref{integral:in:kernel:Chebyshev})). The expllcit form of the
recurrence relation for spin-tomographic star-product kernels is
another new result of the work. This relation provides a
connection of the kernels for qudits ($j \ge 1$) with two basic
kernels for the cases $j=0$ and $j=1/2$. We clarified the
relations between different forms of quantizers and dequantizers
used in spin tomography and available in the literature
\cite{dodonovPLA,oman'ko-jetp,castanos,d'ariano:paini}. We
established that all the different expressions for the quantizers
and dequantizers are equivalent on the set of tomographic symbols
for the spin operators and spin states. The kernel of the dual
tomographic star-product is also expressed in terms of Chebyshev
polynomials (see Eq. (\ref{kernel:dual:Chebyshev})) and calculated
explicitly (see Eq.(\ref{kernel:dual:explicit})). Within the
proposed technique, we also managed to obtain explicit expressions
for delta-function on the tomogram set. In the work
\cite{aniello--ibort-manko}, the relation of irreps characters for
compact and finite groups with kernels of star-products of the
functions on the groups was obtained. In the present work, we
found the relation of the characters of $SU(2)$-group irreps with
the star-product of functions depending on both group element and
weight of irreps.

\begin{acknowledgments}
V.I.M. thanks the Russian Foundation for Basic Research for
partial support under Project Nos. 07-02-00598 and 08-02-90300.
S.N.F. thanks the Ministry of Education and Science of the Russian
Federation and the Federal Education Agency for support under
Project No. 2.1.1/5909.
\end{acknowledgments}

\appendix
\section{\label{appendix:equivalency}Equivalency of star-product kernels}

Since there exist some different explicit forms of the
spin-tomographic star-product kernel, in this Appendix, we
consider the difference $\Delta_{j}$ between the kernel
(\ref{kernel:eplicit:C-G}) expressed in terms of Clebsch-Gordan
coefficients and the kernel (\ref{kernel:explicit:new}) derived on
the basis of the exponential representation of the quantizer and
dequantizer operators. In order to illustrate the deviation
between these kernels one can specify $\Delta_{j} =
\mathcal{K}_{j}''({\bf x}_{3},{\bf x}_{2},{\bf x}_{1}) -
\mathcal{K}_{j}'({\bf x}_{3},{\bf x}_{2},{\bf x}_{1})$ for the
low-spin states. In case of qubits, the reader will have no
difficulty in showing that $\Delta_{j=1/2} = 0$. As far as qutrits
are concerned, the direct computation leads to the following
rather difficult result:

\begin{eqnarray}
&& \Delta_{j=1} = -\frac{1}{36} \Big( 3({\bf n}_2 \cdot {\bf
n}_3)^{2} - 1 \Big) - i \frac{1}{8} m_1 ({\bf n}_2 \cdot {\bf
n}_3) \left({\bf n}_{1} \cdot \left[ {\bf n}_{2} \times {\bf
n}_{3} \right] \right) \nonumber\\
&& + \frac{1}{8} m_1 m_2 \Big( 3 ({\bf n}_{2} \cdot {\bf n}_{3})
({\bf n}_{3} \cdot {\bf n}_{1}) - ({\bf n}_{1} \cdot {\bf n}_{2})
\Big) + \frac{1}{8} m_1 m_3 \Big( 3 ({\bf n}_{1} \cdot {\bf
n}_{2}) ({\bf n}_{2} \cdot {\bf n}_{3}) - ({\bf n}_{3} \cdot {\bf
n}_{1})
\Big) \nonumber\\
&& + \frac{1}{144} (3 m_1^2 - 2) \Big\{ \Big(5 - 3 \left( ({\bf
n}_1 \cdot {\bf n}_2)^2 + ({\bf n}_2 \cdot {\bf n}_3) ^2 + ({\bf
n}_3 \cdot {\bf n}_1)^2 \right) - 9 \left({\bf n}_{1} \cdot \left[
{\bf n}_{2} \times {\bf n}_{3} \right] \right)^2 \Big) \nonumber\\
&& \qquad \qquad + 4 \Big( 3({\bf n}_{1} \cdot {\bf n}_{2})^{2} -
1 \Big) + 4 \Big( 3({\bf n}_{3} \cdot {\bf n}_{1})^{2} - 1 \Big)
\Big\} \nonumber\\
&& + \frac{1}{36} (3m_2^2 - 2) \Big( 5 \left( 3({\bf n}_{1} \cdot
{\bf n}_{2})^{2} - 1 \right) + 2 \Big) + \frac{1}{36} (3m_3^2 - 2)
\Big( 5 \left( 3({\bf n}_{1} \cdot {\bf n}_{2})^{2} - 1 \right) +
2 \Big) \nonumber\\
&& - i \frac{5}{8} m_1 (3 m_2^2 - 2) ({\bf n}_2 \cdot {\bf n}_3)
\left({\bf n}_{1} \cdot \left[ {\bf n}_{2} \times {\bf n}_{3}
\right] \right) - i \frac{5}{8} m_1 (3 m_3^2 - 2) ({\bf n}_2 \cdot
{\bf n}_3) \left({\bf n}_{1} \cdot \left[ {\bf n}_{2} \times {\bf
n}_{3} \right] \right) \nonumber
\end{eqnarray}

\begin{eqnarray}
&& - i \frac{3}{8} (3 m_1^2 - 2) m_2 ({\bf n}_3 \cdot {\bf n}_1)
\left({\bf n}_{1} \cdot \left[ {\bf n}_{2} \times {\bf n}_{3}
\right] \right) - i \frac{3}{8} (3 m_1^2 - 2) m_3 ({\bf n}_1 \cdot
{\bf n}_2) \left({\bf n}_{1} \cdot \left[ {\bf n}_{2} \times {\bf
n}_{3} \right] \right) \nonumber\\
&& + \frac{1}{4} m_1 m_2 (3 m_3^2 - 2) ({\bf n}_1 \cdot {\bf n}_2)
+ \frac{1}{4} m_1 (3 m_2^2 - 2) m_3 ({\bf n}_3 \cdot {\bf n}_1) +
\frac{1}{36} (3 m_2^2 - 2) (3 m_3^2 - 2) \nonumber\\
&& + \frac{1}{144} (3 m_1^2 - 2) (3 m_2^2 - 2) \Big\{ 2 \Big(
3({\bf n}_{3} \cdot {\bf n}_{1})^{2} - 1 \Big) \nonumber\\
&& \qquad \qquad + 5 \Big(5 - 3 \left( ({\bf n}_1 \cdot {\bf
n}_2)^2 + ({\bf n}_2 \cdot {\bf n}_3) ^2 + ({\bf n}_3 \cdot {\bf
n}_1)^2 \right) - 9 \left({\bf n}_{1} \cdot \left[ {\bf n}_{2}
\times {\bf n}_{3} \right] \right)^2 \Big) \Big\} \nonumber\\
&& + \frac{1}{144} (3 m_1^2 - 2) (3 m_3^2 - 2) \Big\{ 2 \Big(
3({\bf n}_{1} \cdot {\bf n}_{2})^{2} - 1 \Big) \nonumber\\
&& \qquad \qquad + 5 \Big(5 - 3 \left( ({\bf n}_1 \cdot {\bf
n}_2)^2 + ({\bf n}_2 \cdot {\bf n}_3) ^2 + ({\bf n}_3 \cdot {\bf
n}_1)^2 \right) - 9 \left({\bf n}_{1} \cdot \left[ {\bf n}_{2}
\times {\bf n}_{3} \right] \right)^2 \Big) \Big\} \nonumber\\
&& + \frac{5}{72} (3 m_1^2 - 2) (3 m_2^2 - 2) (3 m_3^2 - 2) \Big(
\left( 3({\bf n}_{1} \cdot {\bf n}_{2})^{2} - 1 \right) + \left(
3({\bf n}_{3} \cdot {\bf n}_{1})^{2} - 1 \right) \Big).
\end{eqnarray}

\noindent The difference is especially written in a form such that
each summand gives zero while being integrated with tomographic
symbols. The difference of this type is ascribed to the appearance
of redundant terms in the quantizer operator. For instance, in
case of qutrits, the exponential representation of the quantizer
operator contains two additional terms as compared with the
quantizer found in \cite{castanos,filipp-spin-tomography}:

\begin{equation}
\Delta\hat{D}_{j=1}(m, {\bf n}) = \frac{3m^2-2}{6} \left(%
\begin{array}{ccc}
  1 & 0 & 0 \\
  0 & 1 & 0 \\
  0 & 0 & 1 \\
\end{array}%
\right)
+ \frac{1}{6} \hat{R}({\bf n}) \left(%
\begin{array}{ccc}
  1 & 0 & 0 \\
  0 & -2 & 0 \\
  0 & 0 & 1 \\
\end{array}%
\right) \hat{R}^{\dag}({\bf n}).
\end{equation}

\noindent Let us remark that the quantizer enables to reconstruct
the density operator if the state tomogram is given. It can be
easily checked that the integration of the difference
$\Delta\hat{D}_{j=1}(m, {\bf n})$ with any spin tomogram
$w_{j=1}(m, {\bf n})$ gives zero and does not change the density
operator $\hat{\rho}$.

\section{\label{appendix:generalization:of:star:product}Generalization to other tomographic kernels}

The developed approach for calculating the spin-tomographic
star-product kernel can be generalized to other tomographic
kernels. Using the results obtained, one can present universal
formulas which yield all desired kernels, in particular, the
ordinary star-product kernel as well as the dual one and the
expression for delta-function on the set of tomograms.

First we introduce the universal constituent part of the form

\begin{eqnarray}
\label{universal:triple}
&& T_{j}({\bf x}_{3},{\bf x}_{2},{\bf x}_{1},s_3,s_2,s_1) \nonumber\\
&& = \frac{1}{(1-3s_{3}^{2})(1-3s_{2}^{2})(1-3s_{1}^{2})}
\sum\limits_{k=0}^{[j]} \sum\limits_{p_{1}+\dots+p_{8}=2j-2k}
\frac{(-1)^{k} (2j-k)! }{2^{-p_1+p_5+p_6+p_7+2p_8}
k! \prod_{l=1}^{8}p_{l}! } \nonumber\\
&& \qquad \times \Big(1+({\bf n}_{2} \cdot {\bf n}_{3})\Big)^{p_5}
\Big(1+({\bf n}_{3} \cdot {\bf n}_{1})\Big)^{p_6} \Big(1+({\bf
n}_{1} \cdot {\bf n}_{2})\Big)^{p_7} \nonumber\\
&& \qquad \times \Big(1 +({\bf n}_{1} \cdot {\bf n}_{2}) +({\bf
n}_{2} \cdot {\bf n}_{3}) +({\bf n}_{3} \cdot {\bf n}_{1}) - i
\left({\bf n}_{1} \cdot \left[ {\bf n}_{2} \times {\bf n}_{3}
\right] \right) \Big)^{p_8} \nonumber\\
&& \qquad \times \left(%
\begin{array}{c}
  -j-m_1-s_1+k-1 \\
  j-m_1-s_1-k-p_1-p_2-p_4-p_5 \\
\end{array}%
\right) \left(%
\begin{array}{c}
  -j-m_2-s_2+k-1 \\
  j-m_2-s_2-k-p_1-p_2-p_3-p_6 \\
\end{array}%
\right) \nonumber
\end{eqnarray}
\begin{eqnarray}
&& \qquad \times \left(%
\begin{array}{c}
  -j-m_3-s_3+k-1 \\
  j-m_3-s_3-k-p_1-p_3-p_4-p_7 \\
\end{array}%
\right).
\end{eqnarray}

\noindent Then for the ordinary star-product kernel we have

\begin{equation}
K_{j}({\bf x}_{3},{\bf x}_{2},{\bf x}_{1}) = {\rm Tr} \left(
\hat{D}_{j}({\bf x}_{3})\hat{D}_{j}({\bf x}_{2})\hat{U}_{j}({\bf
x}_{1}) \right) = (2j+1)^{2} \sum\limits_{s_{3}=-1}^{1}
\sum\limits_{s_{2}=-1}^{1} T_{j}({\bf x}_{3},{\bf x}_{2},{\bf
x}_{1},s_3,s_2,s_1=0),
\end{equation}

\noindent while the dual star-product kernel reads

\begin{equation}
\label{kernel:dual:explicit} K_{j}^{d}({\bf x}_{3},{\bf
x}_{2},{\bf x}_{1}) = {\rm Tr} \left( \hat{U}_{j}({\bf
x}_{3})\hat{U}_{j}({\bf x}_{2})\hat{D}_{j}({\bf x}_{1}) \right) =
(2j+1) \sum\limits_{s_{1}=-1}^{1} T_{j}({\bf x}_{3},{\bf
x}_{2},{\bf x}_{1},s_3=0,s_2=0,s_1).
\end{equation}

Now we present the universal function for kernels which depend on
two sets of variables ${\bf x}_2$ and ${\bf x}_1$:

\begin{eqnarray}
\label{universal:double} && Q_{j}({\bf x}_{2},{\bf x}_{1},s_2,s_1)
= \frac{1}{(1-3s_{2}^{2})(1-3s_{1}^{2})} \sum\limits_{k=0}^{[j]}
\sum\limits_{p_{1}+\dots+p_{4}=2j-2k} \frac{(-1)^{k} (2j-k)!
}{2^{-p_1+p_4}
k! \prod_{l=1}^{4}p_{l}! } %\nonumber\\
%&& \qquad \times
\Big(1+({\bf
n}_{1} \cdot {\bf n}_{2})\Big)^{p_4} \nonumber\\
&& \qquad \times \left(%
\begin{array}{c}
  -j-m_1-s_1+k-1 \\
  j-m_1-s_1-k-p_1-p_2 \\
\end{array}%
\right) \left(%
\begin{array}{c}
  -j-m_2-s_2+k-1 \\
  j-m_2-s_2-k-p_1-p_3 \\
\end{array}%
\right).
\end{eqnarray}

\noindent Note that this function can be obtained from
(\ref{universal:triple}) if we leave out the third binomial
coefficient, put $p_2=p_5=p_6=p_8=0$, and redesignate $p_4
\rightarrow p_2$.

From (\ref{universal:double}) it is readily seen that the kernel
of the unity operator on the set of spin tomograms reads

\begin{equation}
K_{j}^{\delta}({\bf x}_{2},{\bf x}_{1}) = {\rm Tr} \left(
\hat{D}_{j}({\bf x}_{2})\hat{U}_{j}({\bf x}_{1}) \right) = (2j+1)
\sum\limits_{s_{2}=-1}^{1} Q_{j}({\bf x}_{2},{\bf
x}_{1},s_2,s_1=0).
\end{equation}

Let us now consider the transition from the ordinary tomographic
symbols to the dual ones. The relation between symbols has the
form

\begin{equation}
f_{\hat{A}}^{d}({\bf x}_{1}) = \int f_{\hat{A}}({\bf x}_{2})
K_{j}^{o \rightarrow d}({\bf x}_{2},{\bf x}_{1}) d{\bf x}_{2},
\end{equation}

\noindent where the intertwining kernel reads

\begin{equation}
K_{j}^{o \rightarrow d}({\bf x}_{2},{\bf x}_{1}) = {\rm Tr} \left(
\hat{D}_{j}({\bf x}_{2})\hat{D}_{j}({\bf x}_{1}) \right) =
(2j+1)^{2} \sum\limits_{s_{2}=-1}^{1} \sum\limits_{s_{1}=-1}^{1}
Q_{j}({\bf x}_{2},{\bf x}_{1},s_2,s_1).
\end{equation}

Similarly, a transition from the dual tomographic symbols to the
ordinary ones is defined through

\begin{equation}
f_{\hat{A}}({\bf x}_{1}) = \int f_{\hat{A}}^{d}({\bf x}_{2})
K_{j}^{d \rightarrow o}({\bf x}_{2},{\bf x}_{1}) d{\bf x}_{2},
\end{equation}

\noindent where the intertwining kernel reads

\begin{equation}
K_{j}^{d \rightarrow o}({\bf x}_{2},{\bf x}_{1}) = {\rm Tr} \left(
\hat{U}_{j}({\bf x}_{2})\hat{U}_{j}({\bf x}_{1}) \right) =
Q_{j}({\bf x}_{2},{\bf x}_{1},s_2=0,s_1=0).
\end{equation}

\end{document}